\documentstyle[iopconf1,epsf,psfig]{article}
\begin{document}
\hyphenation{Arbor}
\def\alt{\stackrel{<}{\sim}}
\def\agt{\stackrel{>}{\sim}}
\def\etmiss{E\llap/_T}

\title{
SUSY Dark Matter: Direct Searches vs. Collider Experiments\footnote{Invited 
talk at Second International Conference on Dark Matter in Astrophysics and 
Particle Physics (DARK-98), Heidelberg, Germany, July  20-25, 1998}
}

\author{Michal Brhlik\footnote{E-mail:mbrhlik@umich.edu}}

\affil{Randall Physics Laboratory, University of Michigan, Ann~Arbor,
MI~48109-1120, USA}

\beginabstract
The lightest neutralino in supersymmetric models with conserved R-parity
is an attractive candidate for non-luminous matter in the universe.
If relic neutralinos are indeed present as dark matter in our galaxy, they 
can be directly detected in scattering experiments. This could serve as
an independent search channel for supersymmetry complementary to collider 
experiments. I compare the sensitivity of direct detection experiments 
with the reach for supersymmetry at collider facilities in the framework    
of the minimal supergravity model. 

\endabstract

\section{Introduction}

Low energy supersymmetry  (SUSY) is one of the possible ways nature may have 
chosen to avoid the well-known problem of naturalness occuring in the Higgs 
sector of the Standard Model (SM) of particle physics. The simplest 
supersymmetrized extension of the Standard Model, the Minimal Supersymmetric 
Standard model (MSSM) \cite{MSSM}, introduces superpartners to all SM particles 
in order to control ultraviolet properties of the theory. Global supersymmetry 
is subsequently explicitly broken by a collection of soft-breaking terms 
preserving the UV behavior of the model. In addition, the MSSM conserves 
R-parity, a symmetry
introducing a multiplicative R quantum number equal to $+1$ for ordinary 
particles and $-1$ for their superpartners. This symmetry of the MSSM 
Lagrangian then ensures that the lightest supersymmetric particle (LSP) is 
stable. In order to avoid easy detection, the LSP has to be charge and color 
neutral, and interacts only weakly with the rest of the particle spectrum.
The lightest neutralino $\widetilde Z_1$ is strongly favored to be the LSP in 
the MSSM and, due to its favorable properties, it is also an excellent cold 
dark matter candidate. If indeed some or all of dark matter consists of relic 
neutralinos decoupled from the thermal equilibrium with other particles in the 
early stages of the universe, their direct detection could provide us with an 
independent signal for supersymmetry complementary to standard collider 
experiment searches.

In order to illustrate the general situation, it is useful to consider the 
specific framework of the minimal supergravity (mSUGRA) model \cite{SUGRA}. 
This model assumes that the MSSM and its spectrum arises as a low energy 
effective theory  from  $N=1$ supergravity. SUSY is broken in the 
hidden sector
of the model at energy scale $M\sim 10^{10} {\ \rm GeV}$ and the 
information of SUSY breaking is then gravitationally transmitted to the 
observable sector inducing soft SUSY breaking terms at the $M_{GUT}$ scale. 
These soft breaking terms are typically in the range of a few hundred GeV up to 
a TeV and in the minimal version of the model they consist of a universal set 
of GUT scale parameters: a common scalar mass $m_0$, a common trilinear and 
bilinear couplings $A_0$ and $B_0$, and a common gaugino mass $m_{1/2}$. All 
the soft breaking parameters are evolved from the GUT scale down to the 
electroweak scale by solving the relevant set of coupled renormalization group 
equations and the particle spectrum is calculated. The electroweak symmetry 
is broken radiatively as one of the Higgs mass parameters is driven negative in 
the process of renormalization group evolution and as a consequence the B 
parameter together with the absolute value of the Higgs parameter in the 
superpotential $|\mu|$ can be traded for the ratio of the vacuum expectation 
values of the two neutral Higgs fields $\tan\beta$ and $M_Z$. The resulting 
parameter set
\begin{eqnarray*}
m_0, m_{1/2}, A_0, \tan\beta, sgn(\mu)
\end{eqnarray*}
substantially reduces the number of supersymmetric parameters.
     
\section{Direct detection of neutralinos}

The relic abundance of neutralinos in mSUGRA models depends on the particular 
set of parameters and the superpartner spectrum \cite{bb}. It can be very small 
with $\Omega h^2<10^{-3}$ especially if the total mass of two neutralinos is 
close to the mass of $Z^0$ or the light neutral Higgs boson so that an 
$s$-channel resonance occurs in the neutralino annihilation cross section 
effectively wiping out the relic density. Generally, however, there tends to be 
a significant region of parameter space for relatively small values of $m_0$ 
and $m_{1/2}$ 
where $\Omega h^2$ lies within the cosmologically interesting interval between 
0.01 and 1. Heavy spectra, with the exception of the resonance cases, tend to 
produce values of $\Omega h^2>1$ which would lead to universe less than 10 
billion years old. 
\begin{figure}
%\postscript{fd.ps}{0.3}
\setlength{\epsfxsize}{4.1in}
\setlength{\epsfysize}{4.1in}
\centerline{\epsfbox{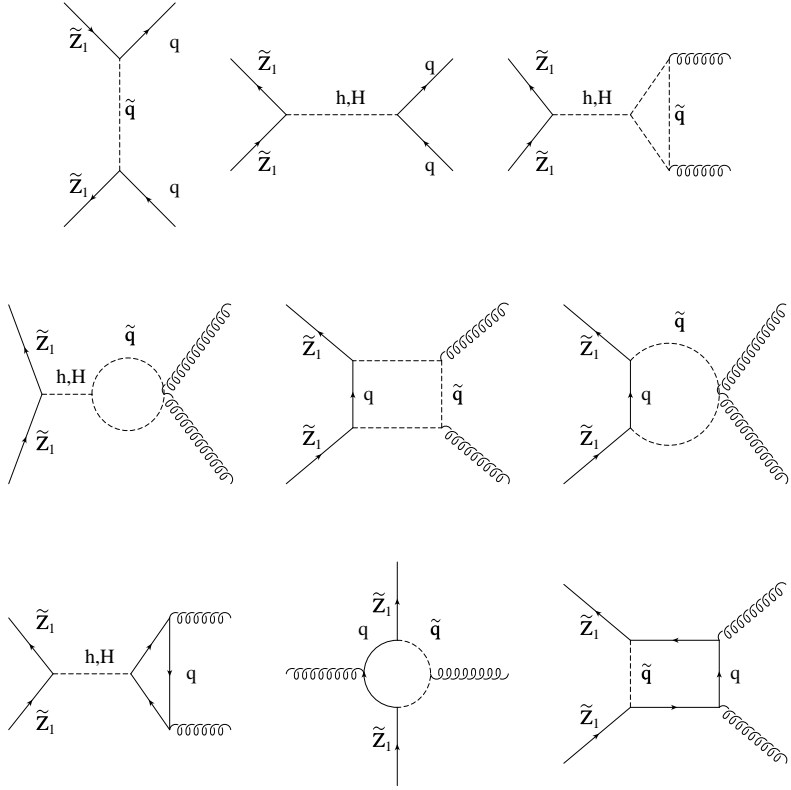}}
\caption{
Feynman diagrams contributing to the scalar part of the effective Lagrangian.
}
\label{fig1}
\end{figure}  
The best way to confirm the existence of ambient relic neutralinos, at least in
our vicinity, would be to devise a detecting method capable of providing a
recognizable experimental signature. The most straightforward method of direct 
detection relies on direct measurement of the neutralino scattering off the 
nuclei in a detector. The main ingredients in the 
calculation of direct detection rates include the neutralino-quark elastic
scattering amplitude, related by the crossing symmetry to the $\widetilde Z_1
\widetilde Z_1\rightarrow q\bar{q}$ annihilation process contributing to the
relic density calculation,  and the amplitude for neutralino scattering on 
gluons proceeding through
one-loop diagrams. The parton level amplitudes then have to be properly
convoluted with quark and gluon distribution functions in nucleons and a
nuclear model must be specified for a particular detector nucleus to
account for structure effects \cite{direc}.

The effective elastic scattering Lagrangian can generally be divided into two 
parts
\begin{eqnarray}
{\cal L}^{eff}_{elastic}={\cal L}^{eff}_{scalar}+{\cal L}^{eff}_{spin}.
\end{eqnarray}
The scalar Lagrangian receives contributions from
neutralino-quark interaction via squarks and Higgs bosons exchange, and from
neutralino-gluon interactions at one-loop level involving quarks, squarks and
the Higgses in the loop diagrams. The parton level Lagrangian includes all 
effective couplings obtained from the matching between the full MSSM and the 
effective theory at scale $Q$ (typically $\sim m_{\widetilde Z_1}$) and can be 
converted into an effective neutralino-nucleon Lagrangian
\begin{eqnarray}
{\cal L}^{eff}_{scalar}&=&f_p \bar{\chi}\chi \bar{\Psi}_p \Psi_p+
f_n \bar{\chi}\chi \bar{\Psi}_n \Psi_n.
\end{eqnarray} 
Effective couplings $f_p$ and $f_n$ contain all the relevant information about 
physics above scale $Q$ and about nucleonic parton structure \cite{dark,bbdet}.      
The differential cross section for a neutralino
scattering off a nucleus $X_{Z}^{A}$ with mass $m_A$ is then expressed as 
\begin{eqnarray}
\frac{d\sigma^{scalar}}{d|\vec{q}|^2}=\frac{1}{\pi v^2}[Z f_p +(A-Z) f_n]^2 
F^2 (Q_r),
\end{eqnarray}                                 
where $\vec{q}=\frac{m_A m_{\widetilde Z_1}}{m_A+m_{\widetilde Z_1}}\vec{v}$ is
the transferred momentum, $Q_r=\frac{|\vec{q}|^2}{2m_A}$ and $F^2(Q_r)$ is the
scalar nuclear form factor. It is obvious from Eqn. (3) that the scalar 
interaction cross section increases quadratically with increasing mass of the 
target nucleus.

%\vskip -1in
Analogously, interaction between the neutralino and nucleon spins is described 
by a spin-dependent nucleon level Lagrangian
\begin{eqnarray}
{\cal L}^{eff}_{spin}&=&2\sqrt{2} (\bar{\chi}\gamma^{\mu} \gamma_5\chi 
\bar{\Psi}_p s_{\mu}\Psi_p+
a_n \bar{\chi}\gamma^{\mu} \gamma_5\chi \bar{\Psi}_n s_{\mu} \Psi_n) ,
\end{eqnarray}  
explicitly involving the nucleon spin vectors $s_{\mu}$.
Coefficients $a_p $ and $a_n$ depend on the nucleonic spin factors 
parametrizing the nucleonic spin matrix elements and on the MSSM parameters 
entering through the parton level effective couplings \cite{dark}. 
For a nucleus with total
angular momentum $J$, the spin interaction differential cross section 
takes the form   
\begin{eqnarray}
\frac{d\sigma^{spin}}{d|\vec{q}|^2}=\frac{8}{\pi v^2}\Lambda^2 J (J+1) 
\frac{S(|\vec{q}|)}{S(0)},
\end{eqnarray}  
where $\frac{S(|\vec{q}|)}{S(0)}$ is the nuclear spin form factor normalized to
1 for pointlike particles, and $\Lambda=\frac{1}{J} [a_p \langle S_p\rangle 
+a_n \langle S_n\rangle]$.
The quantities $\langle S_p \rangle$ and $\langle S_n\rangle$ represent the 
expectation value of the proton (neutron) group spin content in the nucleus.

\begin{figure}
%\postscript{fd.ps}{0.3}
\setlength{\epsfxsize}{2.in}
\setlength{\epsfysize}{1.5in}
\centerline{\epsfbox{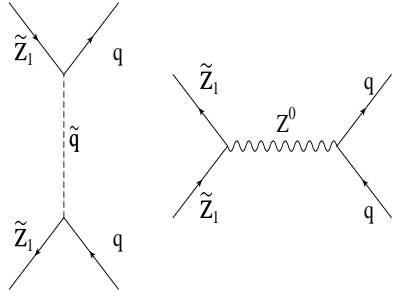}}
\caption{
Feynman diagrams contributing to the spin dependent part of the effective 
Lagrangian.
}
\label{fig2}
\end{figure}

The differential detection rate is calculated by summing the scalar and spin 
interaction contributions and convoluting
them with the local neutralino relic density $\rho_{\widetilde Z_1}$ 
\begin{eqnarray}
\frac{dR}{dQ_r}&=&\frac{4}{\sqrt{\pi^3}}\frac{\rho_{\widetilde Z_1}}
{m_{\widetilde Z_1} v_0} T(Q_r)
\biggl \{ [Z f_p +(A-Z) f_n]^2 F^2 (Q_r)\cr
& &\ \ \ \ \ \ + 8\Lambda^2 J (J+1) \frac{S(|\vec{q}|)}{S(0)} \biggr\},
\end{eqnarray}    
where $v_0\sim 220\,\rm km s^{-1}$ is the circular speed of the Sun around the
center of our galaxy and 
\begin{eqnarray}
T(Q_r)=\frac{\sqrt{\pi}v_0}{2}\int_{v_{min}}^{\infty}
\frac{f_{\widetilde Z_1}(v)}{v}\, dv 
\end{eqnarray}   
integrates over the neutralino velocity distribution. In order to obtain the 
total detection rate, the differential rate in Eqn. (6) has to be integrated
over the relevant interval of $Q_r$ ranging typically from zero up to about
$100\, {\rm keV}$.

It is important to realize that many of the quantities entering the detection 
rate calculation, particularly those related to the nucleonic matrix elements
of the scalar and spin operators, are affected by a fair amount of uncertainty
as they come from experiment. This uncertainty is even bigger for the nucleonic 
spin structure constant. Similarly, the nuclear form factors are subject to the 
assumptions of a particular nuclear model and their reliability depends on the 
degree to which that model accomodates nuclear experimental data \cite{rep}.   
But the most significant source of uncertainty in the detection rate calculation 
is associated with the value of the local neutralino relic density. Here the 
results  depend on particular galactic formation models and the error margin 
can be easily as large as a factor of two or more. It is therefore necessary to 
view 
the numerical results with these limitations in mind.

I will discuss here numerical results obtained for a particular case of a 
$\rm ^{73}Ge$ detector. This isotope is suitable since it is a relatively heavy 
element with mass $m_{Ge}=67.93\,\rm GeV$ and a non-zero total spin 
$J=\frac{9}{2}$. Both the scalar interaction and the spin interaction therefore 
contribute to the elastic scattering cross section. In fact, in some parts of 
the mSUGRA parameter space, particularly for $\mu<0$ and moderate values of 
$\tan\beta$ ($\tan\beta\sim 4-10$), the spin dependent contribution is 
quite comparable in magnitude with the coherent scalar part. This goes 
against the standard view that for sufficiently heavy nuclei the spin dependent 
component is always significantly smaller than the scalar part.  
Another reason for choosing $\rm ^{73}Ge$ is the fact that the nuclear 
properties of this isotope have been 
studied and model calculations of the necessary form factors and expectation 
values are available in the literature \cite{bbdet}.

Qualitative behavior of the detection rate as a function of the mSUGRA 
parameters can be understood based on an analysis of the general formulas 
tracking down the processes contributing most to the scattering cross section 
within the particular framework of minimal supergravity. First of all, Eqn. (6) 
suggests that the detection rate is inversely proportional to the lightest 
neutralino mass $m_{\widetilde Z_1}$. Since the lightest neutralino is almost a 
pure bino (albeit with non-vanishing wino and higgsino admixtures), its mass 
scales with the universal gaugino mass parameter $m_{1/2}$ and so the rate is a 
decreasing function of $m_{1/2}$. It turns out that the dominant part of the 
neutralino-nucleus scattering amplitude comes from the {\it t}-channel heavy Higgs 
boson exchange between the quarks and neutralinos. The Higgs  exchange 
contribution to the $f_N$ coupling is    

\begin{eqnarray}
f_N^{(H)}&=&m_N \sum_{\scriptstyle q=u,d,s} f_{Tq}^{(N)} 
\sum_{j=1,2} \frac{c_{\widetilde{Z}}^{(j)} c_{q}^{(j)}}
{m_{H_j}^2},
\end{eqnarray}
where $f_{Tq}^{(N)}$ is a phenomenological constant. The CP-even Higgs 
couplings are
\begin{eqnarray}
c_{\widetilde{Z}}^{(1)}=\frac{1}{2}(g N_{\ell 2}-g' N_{\ell 1})
(N_{\ell 3}\sin \alpha+N_{\ell 4}\cos\alpha)
\end{eqnarray}
for the lighter Higgs and 
\begin{eqnarray}
c_{\widetilde{Z}}^{(2)}=\frac{1}{2}(g N_{\ell 2}-g' N_{\ell 1})
(N_{\ell 4}\sin \alpha-N_{\ell 3}\cos\alpha)
\end{eqnarray}
for the heavier Higgs, where $\alpha$ is the Higgs mixing angle. 
The quark coefficients are evaluated as 
\begin{eqnarray}
c_{q}^{(i)}=\frac{g}{2 m_W} r_q^{(i)}
\end{eqnarray}
with 
\begin{eqnarray}
r_{u}^{(1)}=-\frac{\sin\alpha}{\sin\beta}&,&
r_{u}^{(2)}=-\frac{\cos\alpha}{\sin\beta}
\end{eqnarray}
for the up type quarks and 
\begin{eqnarray}
r_{d}^{(1)}=-\frac{\cos\alpha}{\cos\beta}&,&
r_{d}^{(2)}=\;\; \frac{\sin\alpha}{\cos\beta}
\end{eqnarray}
\begin{figure}
%\postscript{fd.ps}{0.3}
\setlength{\epsfxsize}{3.8in}
\setlength{\epsfysize}{3.8in}
\centerline{\epsfbox{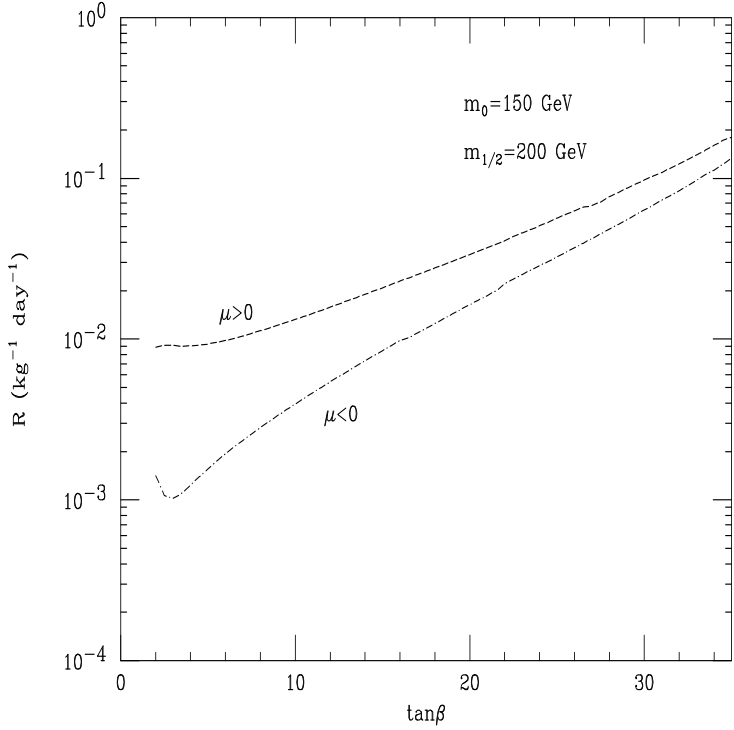}}
\caption{Illustration of the $\rm ^{73}Ge$ detection rate $\tan\beta$ 
dependence for a particular choice of mSUGRA parmaters $m_0=150\,{\rm GeV}$, 
$m_1/2=200\,{\rm GeV}$ and $A_0=150\,{\rm GeV}$, and both signs of $\mu$.
}
\label{fig3}
\end{figure}
for the down type quarks. The contribution from the light Higgs goes through 
zero for $\mu<0$ as the coupling in Eqn. (9) changes sign for $\tan\beta \sim 
3-4$ and the rate for the $\mu<0$ is therefore systematically smaller than for 
$\mu>0$. Moreover, the Higgs couplings to the down type quarks are enhanced by 
a $\tan\beta$ factor and therefore the total detection rate has a tendency to 
grow substantially with increasing $\tan\beta$ as shown in Fig. 3. 
Variation of $A_0$ can also alter the resulting detection rate which generally 
increases with $A_0$ increasing from negative to positive values. It is 
necessary to keep that in mind when analyzing the mSUGRA results.

\section{Collider signals}

Searches for SUSY signals in various channels is one of the major tasks at 
collider experiments. The implications of mSUGRA for both $e^+e^-$ and hadron 
colliders have been studied and experimental reaches for individual channels 
as well as the integrated discovery potential have been worked out. If a signal 
for supersymmetry is discovered, $e^+e^-$ colliders give a better chance of 
disentangling the signal and determining the actual SUSY parameters, especially 
if the electron beam is polarized. The possibility of precise determination of 
the center of mass energy is also an important factor distinguishing the 
$e^+e^-$ colliders from the hadron machines.

For the purposes of comparison between expected SUSY reaches at colliders and 
possible direct detection rates in $\rm ^{73}Ge$, I consider four major 
collider projects, either currently in operation or planned for the near 
future. \newline
{\bf LEP2} is an $e^+e^-$ machine operating at CERN with non-polarized beams. 
The target center of mass energy considered here for SUSY searches was chosen 
at $190\, {\rm GeV}$  while the operating energy for the current run is $189\, 
{\rm GeV}$. In the calculations, an integrated luminosity of $300\, \rm fb^{-1}$ 
was used as a benchmark for the SUSY production processes \cite{LEP}. 
\newline    
{\bf Tevatron} is a $p\bar{p}$ collider operating at $\sqrt{s}=1.8\, \rm TeV $ 
and two major upgrades affecting the SUSY searches are planned for this 
collider. The Main Injector (MI) upgrade will increase the energy to $2\, 
TeV$ and the integrated luminosity will reach $1-2\, \rm fb^{-1}$, so these 
values 
were used in the analysis of MI's potential for discovering SUSY. The 
second upgrade, TeV33, will possibly increase the luminosity up to $25\, 
fb^{-1}$ and will further enhance the reach for SUSY at the Tevatron 
\cite{TeV}.
\newline 
{\bf LHC}, or the Large Hadron Collider, will collide protons with protons 
at the center of mass energy of $14\, {\rm TeV}$. All evaluations of the LHC 
production capabilities in the SUSY channels mentioned in this talk assume an 
integrated luminosity of $10\, fb^{-1}$ \cite{LHC}. 
\newline 
{\bf NLC} - the Next Linear Collider - is a project of a linear $e^+e^-$ 
collider, possibly with 
polarized beams, whose energy at the first stage of operation would be set at 
$500\, {\rm GeV}$ and the luminosity might reach $20\,\rm fb^{-1}$ \cite{NLC}.  

\subsection{$e^+e^-$ collider signals} 
  
{\bf Charginos.} The $e^+e^-\rightarrow {\widetilde W_1}{\widetilde W_1}$ 
chargino production signal occurs in the multi-jet+$\etmiss$ channel, the mixed 
jets+$\ell$+$\etmiss$ channel and the leptonic $\ell$$\bar{\ell^\prime}$+$\etmiss$ 
channel observable both at LEP2 and the NLC. Here and in further text  
electron or muon are meant by leptons. The production usually has a large
cross section and consequently the regions where charginos become observable 
almost coincides with the kinematical limit. Only regions of small $m_0$ 
allowing the decay mode ${\widetilde W_1}\rightarrow 
{\widetilde\nu_{\ell}}\ell$ to turn on producing soft leptons are an exception 
to this rule. The lightest charginos are gauginos and therefore the regions of 
their observability lie below contours of approximately constant $m_{1/2}$.
\newline
{\bf Neutralinos.} Production of the lightest and second lightest neutralino 
$e^+e^-\rightarrow {\widetilde Z_1}{\widetilde Z_2}$ can materialize in the 
detector as a $\ell$$\bar{\ell}$+$\etmiss$ or a jets+$\etmiss$ signal depending 
on the decay mode of $Z_2$. Neutralino signals can in some cases extend beyond 
the reach of the chargino channel since ${\widetilde Z_1}{\widetilde Z_2}$ 
production can be allowed when the production of charginos is disallowed 
because of kinematical constraints. Usually this enhancement occurs in the 
region of smaller $m_0$ and typically is not very big.
\newline
{\bf Sleptons.} In $e^+e^-$ collisions, selectron production is preferred over 
production of other sleptons due to the presence of additional {\it t}-channel 
exchange diagrams. Selectron production results in a clean $e^+e^-$+$\etmiss$
signal with a substantial acollinearity allowing to distinguish it from the 
background SM processes. The selectron signal adds to the SUSY reach mainly in 
the small $m_0$ region where sleptons are light.  

\subsection{Hadron collider signals} 

{\bf Gluinos and squarks.} Production of squarks and gluinos typically results 
in a jets+$\etmiss$ signature although the details of the expected signal 
depend on the particle spectrum and the subsequent decay cascade may also lead 
to the presence of one or more isolated leptons in the event.
\newline
{\bf Chargino-neutralino.} Production of ${\widetilde W_1}{\widetilde Z_2}$ 
pairs in hadronic collisions yields a very strong and distinct signal with
$3\, {\rm leptons}$+$\etmiss$ and no hadronic activity. This signal is 
particularly important for Tevatron SUSY searches where it provides a 
substantial increase of the discovery potential. Additional topologies include   
jets+$3\, {\rm leptons}$+$\etmiss$ and jets+$2\,$opposite sign leptons
+$\etmiss$ where the two leptons come from the neutralino decay and the 
chargino decays hadronically.
\newline
{\bf Sleptons.} The ${\widetilde e^+}{\widetilde e^-}$ production at the LHC 
manifests itself through a clean signal including an opposite sign same flavor 
dilepton+$\etmiss$.

Generally speaking, SUSY signals which are to be expected at hadron colliders 
consist of $m$ jets+$n$ leptons+$\etmiss$ and their potential depends on the 
details of the particular mSUGRA model.

\subsection{Large $\tan\beta$}

The large $\tan\beta$ ($\agt 30$) edge of the mSUGRA parameter space is quite 
different - as far as collider signals are concerned - from the regions with 
low or moderate values of $\tan\beta$. As $\tan\beta$ increases, the $\tau$ 
lepton and bottom quark Yukawa couplings also increase and become comparable to 
the top Yukawa coupling and the gauge couplings. The lightest $\tau$ slepton 
and bottom squark are significantly lighter than their first two generation 
partners. As a result, gluino, chargino and neutralino decays into $\tau$ 
leptons and bottom quarks are enhanced thus obscuring some of the SUSY signals. 
Especially the reach via signatures with isolated leptons is substantially 
limited although some extra sensitivity can be gained by using b-tagging. This 
situation is very important for the Tevatron SUSY reaches \cite{ltTev}
since it greatly reduces (or even wipes out) the SUSY regions reachable in the 
large $\tan\beta$ regime at this machine. The LHC sensitivity, on the other 
hand, is not very much affected. The large production rate helps to overcome 
the decreased number of produced hard isolated leptons and for heavier 
sparticle spectra decay channels into on-shell $W$'s and Higgses open up 
producing hard leptons and ensuring that SUSY will still be observed through 
some of the channels \cite{ltLHC}.

\section{Comparison of detection rates with collider reaches}

The results for $\tan\beta =2$ are summarized in Fig. 3. For all parameter sets 
$A_0$ was set to zero. The SUSY reaches for all four considered experimental 
facilities are shown together with  solid lines indicating the contours of 
constant detection rate in  $\rm ^{73}Ge$.
The region to the right of the contour of $\Omega h^2=1$ is excluded since 
larger values of $\Omega h^2$ require too young a universe.
The reaches for LEP2 and the NLC display the two regions corresponding to the 
chargino signal limit scaling with $m_{1/2}$ and to the slepton signal 
responsible for the bulge appearing at the left edge of the contour. The 
contours do not include the LEP2 and NLC sensitivity to Higgs bosons.    
Frame {\it a} indicates
that in the $\mu <0$ case the largest detection rates favor smaller values of 
$m_{1/2}\alt 150\,\rm GeV$, and $m_0 <200\,\rm GeV$. Even though this
region lies within the reach of both LEP2 and the Tevatron Main Injector
upgrade, the actual rates are $\rm few\,\times 10^{-3}\, events/kg/day $, 
far below any realistic detection experiment sensitivity ($\rm 0.1-1\, 
events/kg/day$). 
It is obvious that the LHC signal from slepton production (and even 
more so the jets+$\etmiss$ signal which covers the whole allowed region of 
parameter space) and the NLC cumulative reach would cover regions with 
extremely small detection rates of the order of $\rm 10^{-4}\, events/kg/day$ 
or less.

The case of $\mu>0$, (shown in frame {\it b}), is different in that the 
region with largest rates predicts rates $\sim 0.1$ events/kg/day, close to the 
current experimental limits. However, regions with relatively large detection
rates also yield small neutralino relic density unless they lie in the
vicinity of a neutralino annihilation pole \cite{bbdet}. 
Therefore, should SUSY be discovered by the LHC in the slepton channel or the 
multi-jet$+E\llap/_T$ channel in the segment of the   
parameter space favoring larger relic density, direct dark matter detection 
experiments would face decreased detection rates. The triangular region of 
Tevatron MI reach for $m_0<200\, {\rm GeV}$ comes mainly from the ${\widetilde 
W_1}{\widetilde Z_2}$ trilepton signal and covers a wide range of relic density 
values.
\begin{figure}
%\postscript{fd.ps}{0.3}
\setlength{\epsfxsize}{3.1in}
\setlength{\epsfysize}{4.6in}
\centerline{\epsfbox{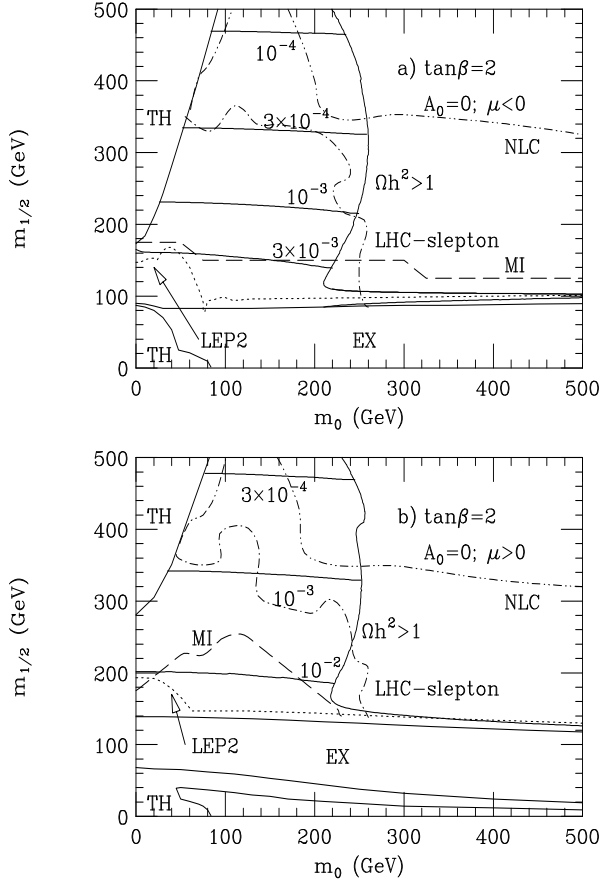}}
\caption{Direct detection rate $R$ in events/kg/day in a $\rm ^{73}Ge$ detector 
for $\tan\beta =2$, a) $\mu <0$ and b) $\mu >0$. Added are SUSY reach contours 
for LEP2, Tevatron MI, LHC slepton signal and NLC.The regions labelled by TH 
are excluded by theoretical considerations while the EX regions are excluded by 
collider searches for SUSY particles. }
\label{fig4}
\end{figure}
The region below the $\rm 10^{-2}\, events/kg/day$ contour could then
possibly be open to competition between the Tevatron MI searches and 
neutralino direct detection experiments with corresponding sensitivity.   

Increasing the value of $\tan\beta$, Fig. 5 displays the results of the 
calculation for $\tan\beta =10$. 
\begin{figure}
%\postscript{fd.ps}{0.3}
\setlength{\epsfxsize}{3.1in}
\setlength{\epsfysize}{4.6in}
\centerline{\epsfbox{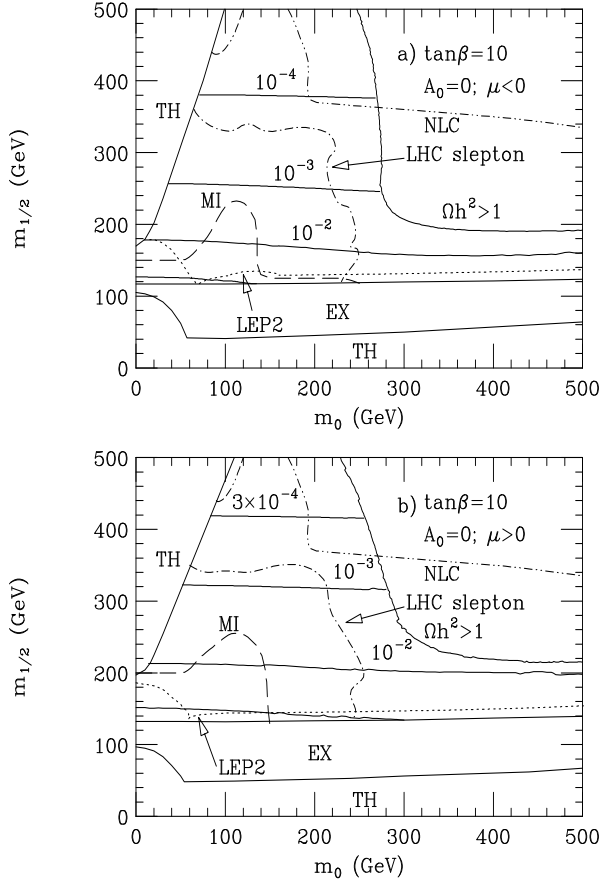}}
\caption{
Same as Fig.4, but for $\tan\beta =10$ 
}
\label{fig5}
\end{figure}
The situation in frame {\it b} is similar to the $\tan\beta =2$ case but one 
has to realize that the detection contours here are shifted towards larger 
values of $m_{\frac{1}{2}}$. For $\mu <0$ we see a significant increase in the 
detection rate. This is due to the fact that the
effective contribution from the Higgs bosons exchange, which dominates the
scalar cross section as quark-squark couplings are suppressed by chiral 
symmetry and large squark masses, increases with $\tan\beta$. 
Also, the Tevatron MI trilepton signal in both frames covers only part of the 
region with detection rates larger than $\rm 10^{-2}\, events/kg/day$. In 
addition to the region where the direct detection experiments will be competing 
against the Tevatron signal there is also a region with $m_0>150\, {\rm GeV}$
where direct detection could be the only chance of observing a supersymmetric 
signal before the LHC comes into operation. Existence of this region is 
mostly due to the presence of the neutralino annihilation poles which increase 
the annihilation cross section and decrease the relic density in their 
vicinity. For this value of $\tan\beta$ too, the LHC and the NLC will cover the 
mSUGRA parameter space up to regions yielding very small deetection rates of 
the order of $10^{-4}\, \rm events/kg/day $.  
\begin{figure}
%\postscript{fd6.ps}{0.3}
\setlength{\epsfxsize}{3.1in}  
\setlength{\epsfysize}{4.6in}
\centerline{\epsfbox{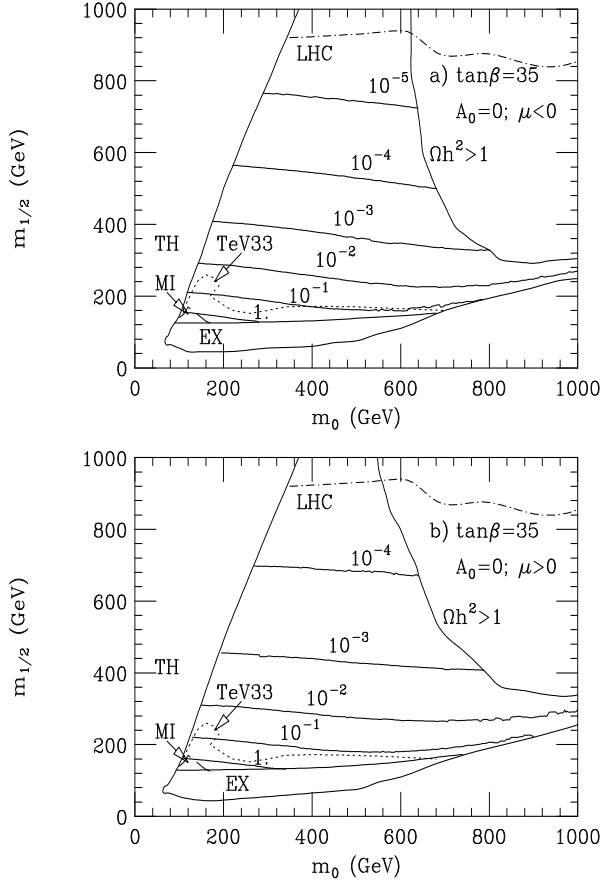}}
\caption{Same as Fig. 4, except for $\tan\beta=35$. Shown are SUSY reach 
contours for the Tevatron MI and TeV33 upgrades, and for the cumulative 
LHC search.}
\label{fig6}
\end{figure}

Finally, Fig. 6 shows the detection rate contours and collider reaches for 
the case of  $\tan\beta=35$. Since the region of cosmologically interesting 
relic density $\Omega h^2$ has grown considerably compared to previous two 
plots the scale in this plot is also expanded. The first thing to be noted is 
the fact that detection rates for smal values of $m_{1/2}$ can be larger than 
$1\,\rm event/kg/day$. Unfortunately, this occurs in the region where the relic 
density is uninterestingly small, not even large enough to provide a suficient 
amount of relic neutralinos to explain the rotation curves of galaxies. 
However, big enough detection rates can be achieved for values of $m_{1/2}$ up 
to $\rm 300\, GeV$  with corresponding relic density values much more 
interesting from the cosmological point of view and reaching up to $\Omega 
h^2\simeq 
0.5$. SUSY discovery potential contours for both upgrades of the Tevatron, the 
Main Injector and TeV33, are plotted. The MI reach is very tiny and covers only 
a part of the region with detection rates above $1\, \rm event/kg/day$. The 
TeV33 
upgrade improves the situation but still large pieces of the parameter space 
available to detectors with sensitivities better than $10^{-2}\, \rm 
event/kg/day$ 
remain out of reach for the Tevatron. 
This is even more interesting since the 
uncovered regions generally yield larger relic densities. Therefore in the 
large $\tan\beta$ regime the direct detection experiments stand a good chance 
of being unchallenged in the search for SUSY until the LHC becomes operational. 
The cumulative reach of the LHC from the jets+$\etmiss$ and 
jets+$\ell$+$\etmiss$ signals will cover cover the rest of the parameter space
where detection rates are too small for direct experiments.    

One remark concerning the available parameter space in Figs. (3)-(6) is due 
here. Various independent constraints, such as from the $b\rightarrow s\gamma$
process \cite{bsg}, from the requirement of no false vacua in the theory 
\cite{vac}, from fine tuning \cite{ft} etc., limit the extent of the available 
parameter space, in some cases quite severely. For the sake of brevity these 
constraints have not been discussed here but have been studied in literature.  
    
\section{Impact of CP violating phases}

The minimal SUGRA model is parametrized in terms of only four parameters plus 
one sign and the sparticle spectrum generated starting from this set is 
therefore highly correlated. By relaxing unification conditions for the scalar 
soft breaking masses the number of free parameters increases while the gaugino 
mass unification (inspired by the gauge coupling unification) can still be 
preserved. These non-universalities are certainly legitimate and their effect 
on the predicted neutralino relic density and detection rate can be studied 
\cite{arno}. 

Another important issue which needs to be addressed in connection with the 
minimal SUGRA parameter set is the question of CP violating phases. The MSSM 
includes a fairly extensive collection of parameters entering the SUSY breaking 
sector and many of them are CP violating complex phases. In the simplest 
extension of the mSUGRA model, the common gaugino mass $m_{1/2}$, the common 
trilinear parameter $A_0$, the Higgs mass parameter in the superpotential $\mu$ 
and the Higgs bilinearterm $B_0$ can all be complex. Redefinition of the Higgs 
fields allows one to eliminate one phase in the Higgs sector and take, for 
instance, $B_0$ to be real. The MSSM possesses one more approximate symmetry, 
namely the R-symmetry, which is violated by the gaugino mass term and can be 
used to rotate away the phase of $m_{1/2}$. As a result, the simplest extension 
of the mSUGRA framework still observing universality introduces two complex 
phases $\varphi_{\mu}$ and $\varphi_{A_0}$. Not all values of the phases are 
allowed by the experimental limits on the electron and 
neutron dipole moments but, as discussed recently \cite{dip}, large values of 
the phase can still be allowed. 

The calculation of $\Omega h^2$ requires a knowledge of the elements of the 
neutralino mass matrix, all of which enter into calculating the neutralino 
annihilation cross section that determines the neutralino relic density. The 
cross sections depends strongly on the amount of gaugino-higgsino mixing and 
the resulting relic density can be very different if phases are allowed as 
parameters in the model. $\Omega h^2$ can vary by a factor of a few depending 
on the phases.
Similarly, the neutralino-nucleon scattering cross section can be very 
sensitive to the presence of complex phases and the limits on neutralino 
detection need to be re-evaluated in the extended framework \cite{phase}.

\section{Conclusion}

The lightest neutralino in supersymmetric models is a very appealing
candidate to provide some or all of the cold dark matter (CDM) of the universe. 
Even in the restrictive framework of the minimal supergravity model with 
radiative electroweak symmetry breaking one can find regions of parameter space 
where the neutralino relic density ranges within the cosmologically relevant 
interval of values. These regions are further restricted by constraints from
experimental limits on rare decays, from the requirement of the absence of  
non-standard global minima in the theory and by other constraints, but 
typically there are still regions left which are consistent with all 
constraints and lead to interesting neutralino relic densities. Most of those 
regions, with the exception of some areas in the vicinity of the neutralino 
annihilation poles, are also favored by fine tuning considerations. 
  
From the experimental point of view it is interesting that some of these 
cosmologically favored regions are within the reach of collider experiments at 
LEP2 and the Tevatron upgrades but certainly most of the relevant parameter 
space will be covered by searches for SUSY at the LHC and the NLC.
Although the collider experiment searches are the main way of looking for 
SUSY signals in nature, direct detection of ambient neutralinos could 
provide a useful tool for SUSY searches. The neutralino detection rate in 
direct scattering experiments grows proportionally to $\tan\beta$ while the 
experimental reach for SUSY at LEP2 and the Tevatron is decreasing due to 
decreasing branching ratios for the electron and muon production. In this 
sense, the direct detection searches are complementary to the collider 
experiments.

Dark matter detectors reaching a sensitivity of $R\sim 10^{-2}\,\rm 
events/kg/day$ would usually have a better reach in the mSUGRA parameter space 
then LEP2 would for SUSY particles (excluding the Higgs boson). It would also 
have a comparable reach with the Tevatron MI or even exceed it in some regions
with intermediate values of $\tan\beta$. In the large $\tan\beta$ region,
the chances of detecting SUSY become much better for direct detection 
experiments and their reach is even better then the expected reach at the Tev33 
Tevatron upgrade. It is therefore possible that in the interesting time period 
before the LHC becomes operational the fist signal for SUSY will come from 
direct detection experimets.

\section*{Acknowledgments}

I would like to thank the organizers of DARK-98 for their invitation to the 
conference and for their kind hospitality.  
I am grateful to Howie Baer with whom much of the work presented in this talk
was done and to Gordy Kane for collaboration and discussions on the topic of CP 
violating phases.

\end{document}